\documentclass[
	a4paper, 
	10pt, 
	twoside, 
]{LTJournalArticle}

\usepackage{amsmath}
\usepackage{nameref}
\usepackage{varioref}
\usepackage{hyperref}
\usepackage{cleveref}
\usepackage[caption = false]{subfig}
\usepackage{subcaption}
\usepackage{multirow}
\usepackage{color}
\runninghead{Flows for unsupervised Anomaly Detection at 40 MHz} 

\title{It's not a FAD:\\ first results in using Flows for unsupervised Anomaly Detection at 40 MHz at the Large Hadron Collider} 

\author{%
	Francesco Vaselli\textsuperscript{1,2}\thanks{Corresponding author: \href{mailto:francesco.vaselli@cern.ch}{francesco.vaselli@cern.ch}}, Chang Sun\textsuperscript{5}, Thea Aarrestad\textsuperscript{4}, Dimitrios Danopoulos\textsuperscript{1},\\ Roope Oskari Niemi\textsuperscript{1}, Maciej Mikolaj Glowacki\textsuperscript{1}, Katya Govorkova\textsuperscript{3},
Vladimir Loncar\textsuperscript{6}, \\ Felice Pantaleo\textsuperscript{1} and Maurizio Pierini\textsuperscript{1}
}

\date{\footnotesize\textsuperscript{\textbf{1}}European Organization for Nuclear Research (CERN) CH-1211 Geneva 23, Switzerland\\\footnotesize\textsuperscript{\textbf{2}}also at Scuola Normale Superiore, Pisa, Italy and Istituto Nazionale di Fisica Nucleare (INFN) sezione di Pisa, Italy\\\footnotesize\textsuperscript{\textbf{3}}Massachusetts Institute of Technology, Cambridge, MA, USA\\
\footnotesize\textsuperscript{\textbf{4}}ETH, Federal Institute of Technology Zurich, Zurich, CH\\
\footnotesize\textsuperscript{\textbf{5}}California Institute of Technology, Pasadena, CA, USA\\
\footnotesize\textsuperscript{\textbf{6}}Institute of Physics, Belgrade, Serbia}

\renewcommand{\maketitlehookd}{%
	\begin{abstract}
		\noindent We present the first implementation of a Continuous Normalizing Flow (CNF) model for unsupervised anomaly detection within the realistic, high-rate environment of the Large Hadron Collider's L1 trigger systems. While CNFs typically define an anomaly score via a probabilistic likelihood, calculating this score requires solving an Ordinary Differential Equation, a procedure too complex for FPGA deployment. To overcome this, we propose a novel, hardware-friendly anomaly score defined as the squared norm of the model's vector field output. This score is based on the intuition that anomalous events require a larger transformation by the flow. Our model, trained via Flow Matching on Standard Model data, is synthesized for an FPGA using the hls4ml and da4ml libraries. We demonstrate that our approach effectively identifies a variety of beyond-the-Standard-Model signatures with performance comparable to existing machine learning-based triggers. The algorithm achieves a latency of a few hundred nanoseconds, or even less when using advanced quantization techniques, and requires minimal FPGA resources, establishing CNFs as a viable new tool for real-time, data-driven discovery at 40 MHz.
	\end{abstract}
}

\newcommand{\LQbtau}{$LQ \to b\tau$}
\newcommand{\Allll}{$A \to 4\ell$}
\newcommand{\htautau}{$h^0 \to \tau\tau$}
\newcommand{\htaunu}{$h^{\pm} \to \tau\nu$}
\newcommand{\model}{\multirow{2}{*}{Model}}
\newcommand{\adscore}{\multirow{2}{*}{AD score}}
\newcommand{\tpr}{\multicolumn{4}{c|}{TPR @ FPR $10^{-5}$ [\%]}}
\newcommand{\auc}{\multicolumn{4}{c}{AUC [\%]}}

\begin{document}

\maketitle 


\section{Introduction} \label{sec1}

The 40 million proton-proton collisions per second at the CERN Large Hadron Collider (LHC) \autocite{Evans_2008} present a formidable data challenge. The sensors in the ATLAS \autocite{ATLAS_Collaboration_2008} and CMS \autocite{CMS_Collaboration_2008} detectors output hundreds of terabytes of data per second, which must be filtered in real-time to decide what to save on disk for further analysis. This is accomplished by a two-stage trigger system: a hardware-based Level-1 Trigger (L1T) \autocite{Sirunyan_2020, CERN-LHCC-2020-004, collaboration_2020, CERN-LHCC-2017-020} on \emph{Field Programmable Gate Arrays} (FPGAs), which reduces the data rate by almost a factor of 1000 within microseconds, followed by a software-based High-Level Trigger (HLT) on a CPU farm.

The selection algorithms in these triggers are typically supervised, targeting specific physics signatures motivated by theory, as exemplified by the successful search for the Higgs boson \autocite{20121, 201230}. This paradigm, however, may fail to identify unforeseen new physics. Consequently, there is significant community effort to use unsupervised machine learning for model-agnostic searches \cite{Aarrestad_2022, Kasieczka_2021}. These efforts have explored autoencoders \autocite{kingma2022autoencodingvariationalbayes} for offline processing and have proposed their integration into the HLT to capture rare events in a special data stream \autocite{Cerri_2019, knapp2020adversariallylearnedanomalydetection}, echoing earlier non-ML strategies \autocite{Francesco:1306501}.

More recently, a new approach to anomaly detection (AD) argued for deploying these algorithms earlier in the data processing chain: at the L1T. Placing an AD search at this stage avoids the selection biases introduced by the standard L1T algorithms, which often discard potentially interesting low-energy topologies. An unbiased, model-independent AD trigger can select events based on their abnormality, not a specific signature, thus maximizing discovery potential. Govorkova et al. \autocite{Govorkova_2022} pioneered this new type of strategy using autoencoders as the AD algorithm to be ported to the L1T.

Since 2023, the CMS Collaboration is operating autoencoder-based anomaly detection algorithms. At first, a VAE based on high-level features (named AXOL1TL \autocite{gandrakota2024realtimeanomalydetectionl1}) was deployed, following the strategy highlighted in Ref \autocite{Govorkova_2022}. Then, a Convolutional Autoencoder (named CICADA \autocite{gandrakota2024realtimeanomalydetectionl1}) was put in use, using Knowledge Distillation as a compression algorithm \autocite{hinton2015distillingknowledgeneuralnetwork}. In 2025, ATLAS took the first steps towards the same strategy, with the deployment of GELATO \autocite{Cohen:2938881}. All these algorithms consider autoencoders as the baseline tool for anomaly detection in real-time, as originally proposed in Ref. \autocite{Govorkova_2022}.

To the best of our knowledge, the present work is the first proof of concept for the application of a new type of ML model at the L1T for unsupervised AD: \emph{Continuous Normalizing Flows} (NF) \autocite{papamakarios2021normalizingflowsprobabilisticmodeling}.

Discrete NF have already been considered as anomaly detection tools for high energy physics, but only in the context of offline data analysis, see \autocite{Kasieczka_2021, Krause_2024, Jawahar_2022}.


Implementing these algorithms at the L1T, with its severe latency (< 1 µs) and resource constraint, is made possible by open-source \texttt{hls4ml} \autocite{fastml_hls4ml, Duarte:2018ite, Aarrestad:2021zos, Loncar:2020hqp} and \texttt{da4ml} \autocite{da4ml} libraries, providing a pathway to deploy neural networks and other ML models on FPGAs. By generating highly optimized, fully on-chip firmware, these tools can meet the L1T latency and throughput requirements (initiation interval < 150 ns, related to the bunch-crossing time for the LHC operations). Moreover, they support quantization-aware training (QAT) \autocite{Coelho_2021} that potentially allows extreme model compression and a further reduction of FPGA resource consumption.

The main contributions of this work are the following:
\begin{itemize}
    \item We train a Continuous NF on a realistic dataset of low-level features of physics objects Standard Model (SM) signatures, and from there we define an \emph{anomaly detection score} suitable for inference on FPGA;
    \item We evaluate results on different benchmarks for new physics scenarios at Floating Point precision, comparing it to a similar architecture of Govorkova et al. \autocite{Govorkova_2022}, and observing similar performances;
    \item We compress the model through different strategies, namely \emph{Post Training Quantization} (PTQ) or \emph{Quantization Aware Training} (QAT) using \emph{High Granularity Quantization} (HGQ). We then port the results to FPGA, showing that the resource usage is well within the requirements of a L1T trigger system.
\end{itemize}

\section{Data Samples} \label{sec2}
We use the datasets employed by \autocite{Govorkova_2022},  published on Zenodo \autocite{thea_aarrestad_2021_5046446, thea_aarrestad_2021_5055454, thea_aarrestad_2021_5061633, thea_aarrestad_2021_5061688} and discussed in detail in \autocite{govorkova2021lhcphysicsdatasetunsupervised}.

The SM data sample represents a typical proton-proton collision dataset that has been pre-filtered by requiring the presence of an electron or a muon with a transverse momentum $p_{\text{T}} > 23\,\text{GeV}$ and a pseudo-rapidity $|\eta| < 3$ (electron) and $|\eta| < 2.1$ (muon).
This is representative of a typical L1T selection algorithm of a multipurpose LHC experiment.
In addition to this, we consider the four benchmark new physics scenarios discussed in Cerri et al. \autocite{Cerri_2019}:
\begin{itemize}
    \item A leptoquark (LQ) with a mass of $80\,\text{GeV}$, decaying to a $b$ quark and a $\tau$ lepton: \LQbtau \;\autocite{thea_aarrestad_2021_5055454},
    \item A neutral scalar boson ($A$) with a mass of $50\,\text{GeV}$, decaying to two off-shell Z bosons, each forced to decay to two leptons: \Allll \;\autocite{thea_aarrestad_2021_5046446},
    \item A scalar boson with a mass of $60\,\text{GeV}$, decaying to two tau leptons: \htautau \;\autocite{thea_aarrestad_2021_5061633},
    \item A charged scalar boson with a mass of $60\,\text{GeV}$, decaying to a tau lepton and a neutrino: \htaunu \;\autocite{thea_aarrestad_2021_5061688}.
\end{itemize}
These four processes are used to evaluate the accuracy of the trained models.

In total, we use 3.5 million events from the background sample for training. The new physics benchmark samples are only used for evaluating the performance of the models, along with an additional 2 million background events not used in training.

\subsubsection{Preprocessing}
We apply a simple standard-scaling operation in training and inference. This operation is usually replicated on FPGA through a bit-shift, as it involves just subtraction and division, and, while it implies the use of a limited amount of extra resources, it could theoretically be applied upstream in the board and be used for running multiple algorithms. Thus, it is not included in the resources estimate of the flow algorithm.

We note that the performance of the algorithm is very much dependent on the specific preprocessing operation being applied. 


\section{Normalizing Flow Models} \label{sec3}

Normalizing Flows (NFs) are a class of generative models that learn an explicit representation of an unknown data probability density function (pdf), $p(\mathbf{x})$. They achieve this by defining an invertible and differentiable mapping, $f$, between the complex data space and a simple base distribution (typically a standard Gaussian), $p(\mathbf{z})$:
\begin{equation}
    \mathbf{x} = f(\mathbf{z}) \quad \text{and} \quad \mathbf{z} = f^{-1}(\mathbf{x}).
\end{equation}
We refer to \autocite{papamakarios2021normalizingflowsprobabilisticmodeling} for a comprehensive review of existing NF algorithms.

\subsection{Continuous Normalizing Flows}
A \emph{Continuous Normalizing Flow} (CNF) defines the invertible mapping $f$ not as a single function, but as the solution to a differential equation parameterized by a continuous variable $t \in [0, 1]$. This establishes a "probability path" that smoothly transforms the base distribution $p_0$ into the target data distribution $p_1$. The dynamics of this transformation are governed by a time-dependent vector field $\mathbf{v}_t$, which is parameterized by a neural network with parameters $\phi$:
\begin{equation}
    \frac{d\mathbf{z}(t)}{dt} = \mathbf{v}_t(\mathbf{z}(t), t | \phi),
    \label{eq:cnf_ode}
\end{equation}
with the initial condition $\mathbf{z}(0)$ being a sample from the base distribution, $\mathbf{z}(0) \sim p(\mathbf{z})$. A data-like sample $\mathbf{x}$ is generated by solving this \emph{Ordinary Differential Equation} (ODE) from $t=0$ to $t=1$, yielding $\mathbf{x} = \mathbf{z}(1)$. Note that the dimensionality of the vector field $\mathbf{v}_t$ is the same as the input features $\mathbf{x}$.

\subsection{Flow Matching}
A powerful and stable method for training CNFs is \emph{Flow Matching}~\cite{lipman2023flowmatchinggenerativemodeling, lipman2024flowmatchingguidecode}. This approach recasts the training objective as a simple regression problem. The goal is to make the model's vector field $\mathbf{v}_t$ match a predefined target vector field $\mathbf{u}_t$. This target field $\mathbf{u}_t$ is constructed to transport samples from the base distribution towards samples from the data distribution along a specific "probability path" $p_t$.

Crucially, the path $p_t$ and its corresponding vector field $\mathbf{u}_t$ can be constructed conditionally for each training sample $\mathbf{x}$. This simplifies the training into a regression of the model's output $\mathbf{v}_t$ onto the target $\mathbf{u}_t$ at random points along these paths. The loss function for the model parameters $\phi$ is then:
\begin{equation}
    \mathcal{L}_{\text{FM}}(\phi) = \mathbb{E}_{t, \mathbf{x}, \mathbf{z}_t} \left[ \|\mathbf{v}_t(\mathbf{z}_t, t | \phi) - \mathbf{u}_t(\mathbf{z}_t, t | \mathbf{x})\|^2 \right],
\end{equation}\label{eq:loss}

where $t \sim \mathrm{Unif}[0,1]$, $\mathbf{x} \sim p(\mathbf{x})$, and $\mathbf{z}_t$ is a point sampled from the conditional path distribution $p_t(\mathbf{z}|\mathbf{x})$. This is a straightforward regression loss, making the model relatively easy to train. In this work, we draw from recent developments reviewed in \cite{lipman2024flowmatchingguidecode}. We refer the reader there for a detailed discussion on the construction of various probability paths and their associated vector fields.

\subsection{Model architecture and training}

The input to our model consists of the kinematic variables---transverse momentum ($p_{\text{T}}$), pseudorapidity ($\eta$), and azimuthal angle ($\phi$)---for 18 reconstructed physics objects. These objects are ordered by type: 4 muons, 4 electrons, and 10 jets. The list is augmented with the missing transverse energy (MET), for which the magnitude and $\phi$ are used, while its $\eta$ component is set to zero. This creates a fixed-size input tensor of shape (19, 3). For events with fewer than the maximum number of objects of a given type, the corresponding input slots are zero-padded, a standard practice in L1T algorithm design.

Before being processed by the network, this (19, 3) tensor is flattened into a single 57-dimensional vector. As detailed previously, a standard scaling transformation is applied to these input features. This preprocessing step is a simple affine transformation that can be readily implemented on an FPGA as a bit-shift operation; since its resource usage is minimal, it is omitted from the following discussion.

The core of our model is the neural network that parameterizes the time-dependent vector field $\mathbf{v}_t(\mathbf{z}_t, t | \phi)$ of the Continuous Normalizing Flow. After a basic hyperparameter scan, we settle for simple multilayer perceptron (MLP) architecture consisting of two hidden layers, each with 16 nodes, and using the Rectified Linear Unit (ReLU) as the activation function. The network takes a 58-dimensional flattened input vector, i.e. the 57 physics features plus an additional input for the current timestamp \emph{t}, and outputs a 57-dimensional vector, which represents the vector field $\mathbf{v}_t$ required for the flow transformation. This simple architecture amounts to 1,913 trainable parameters.

The model was trained for 100 epochs using the Adam optimizer with a learning rate of $10^{-3}$ and a batch size of 1024. The network parameters were optimized by minimizing the Flow Matching loss described in section~\ref{sec3} (with the scheduler called \texttt{CondOTScheduler()} and the \texttt{AffineProbPath()} defined in \autocite{lipman2024flowmatchingguidecode} to build the loss function), using the \texttt{PyTorch} \autocite{paszke2019pytorchimperativestylehighperformance} and \texttt{flow-matching} \autocite{lipman2024flowmatchingguidecode} libraries.

\subsection{VAE model for comparison}

In order to compare to the fully-connected vae architecture of Ref. \autocite{Govorkova_2022}, we decide to retrain the same DNN VAE architecture. We refer the reader to section 3 of that work for details about the implementation; we train with the same conditions and the hyperparameter $\beta = 0.8$, plus the same preprocessing used for the flow architecture.

We could not exactly reproduce the results obtained in Ref. \autocite{Govorkova_2022}. In addition to possible differences in the underlying training software, a possible cause of discrepancy could originate from undocumented pre-processing steps, with respect to the code published by the authors, see \autocite{codeVAE1, codeVAE2}. In our experiments, we do observe a large performance dependence on the applied processing. For consistency reasons, we take our trained version of the VAE as the reference in our plots, but we still report the original results of \autocite{Govorkova_2022} in the summary tables. 

\section{Anomaly Detection Scores}\label{sec4}

Once trained, Continuous Normalizing Flows map an input data point $\mathbf{x}$ to a latent variable $\mathbf{z}$ by solving the ODE that defines the model, see Eq. \ref{eq:cnf_ode}.

We denote as $\mathbf{v}_t(\mathbf{z}(t), t | \phi)$ the neural network with parameters $\phi$ that defines a time-dependent vector field. For this application, the integration runs from $t=1$, where $\mathbf{z}(1) = \mathbf{x}$, to $t=0$, where $\mathbf{z}(0)$ is the corresponding point in the latent Gaussian space.

\subsubsection{Flow ODE}
The canonical approach to define an anomaly score for flows is the negative Gaussian log-likelihood of the latent point $\mathbf{z}(0)$. This score has a direct probabilistic interpretation, making it an elegant definition. However, calculating it requires integrating the ODE, a procedure too complex to implement efficiently on an FPGA. This would involve:
\begin{enumerate}
    \item Starting from the data point $\mathbf{x}$ at timestep $t=1$.
    \item Executing a forward pass of the model to get the vector field $\mathbf{v}_t$.
    \item Solving the current step of the ODE integrator using this vector field.
    \item Repeating steps 2 and 3 until the integration reaches $t=0$ to find the latent point $\mathbf{z}(0)$.
    \item Evaluating the Gaussian log-likelihood of $\mathbf{z}(0)$ to get the final score.
\end{enumerate}
This iterative process is clearly infeasible within the strict latency constraints of the L1T. In what follows, we still compute this score on a CPU with floating-point precision as a comparison, and we refer to it as \emph{Flow ODE}. We show results obtained using the \texttt{euler} solver with just 2 timesteps from 1 to 0. From a preliminary investigation, when increasing the timesteps of integration performances are anyways similar to those showed in \cref{sec:results} for the current choice of timesteps.

\subsubsection{Flow $\mathbf{v}_t$}

To create an FPGA-compatible alternative, we need an anomaly score that avoids this integration. We propose a new score based on the intuition that anomalous points, being further from the training distribution, require a larger "push" from the vector field to be mapped toward the latent prior. Therefore, the magnitude of the vector field itself, i.e. the output of the model, can serve as a proxy for anomaly. We define our score as the squared norm of the vector field evaluated at the initial point $\mathbf{x}$:
\begin{equation}
    \mathcal{AS}(\mathbf{x}) = \|\mathbf{v}_t(\mathbf{x}, t | \phi)\|^2 = \sum_i \left( v_t(\mathbf{x}, t | \phi)_i \right)^2.
    \label{eq:anomaly_score}
\end{equation}
This strategy has the key advantage of requiring only a single forward pass of the model, making it suitable for fast inference. We refer to this approach as \emph{Flow $\mathbf{v}_t$}.

The value of this score depends on the timestep $t$ at which the model is evaluated. For this work, we choose to evaluate the field at the beginning of the trajectory, $t=1$, as this is where the model first encounters the input data. We note that other choices are possible, and recent developments in single-step or straight-path flow models may offer other optimal evaluation points.

\subsubsection{VAE Score}
In order to present a comparison with a similar architecture as the one used in \autocite{Govorkova_2022}, we use the Kullback-Leibler score $D_{KL}$ used in the loss function as the anomaly score for this model:

\begin{equation}
    D_{KL} = \frac{1}{2} \sum_{j=1}^{2} (\sigma_{j}^2 + \mu_{j}^2 - \log(\sigma_{j}^2) - 1)
\end{equation}

and we refer to it as \emph{VAE} or $D_{KL}$ in the following.

\subsubsection{Further considerations and outlook}
Finally, we would like to note that a vector-field-based approach also opens avenues for other potential scores, such as the \emph{divergence} of the field (at the cost of multiple model evaluations), and provides a conceptual link to other ODE-based generative models like \emph{Diffusion Models}~\autocite{ho2020denoisingdiffusionprobabilisticmodels}. In this paper, however, we focus on the first two approaches described, leaving the investigation of these alternatives to future work.

\section[Results]{Results at Floating Point precision}\label{sec:results}
The model performance is assessed using the four new
physics benchmark models. The anomaly detection scores
considered are the \emph{Flow} $\mathbf{v}_t$ and \emph{Flow ODE} (as offline comparison) for the flow model, and the $D_{KL}$ for the VAE model.

\begin{figure*}
    \centering
    \subfloat
    {\includegraphics[width=.5\textwidth]{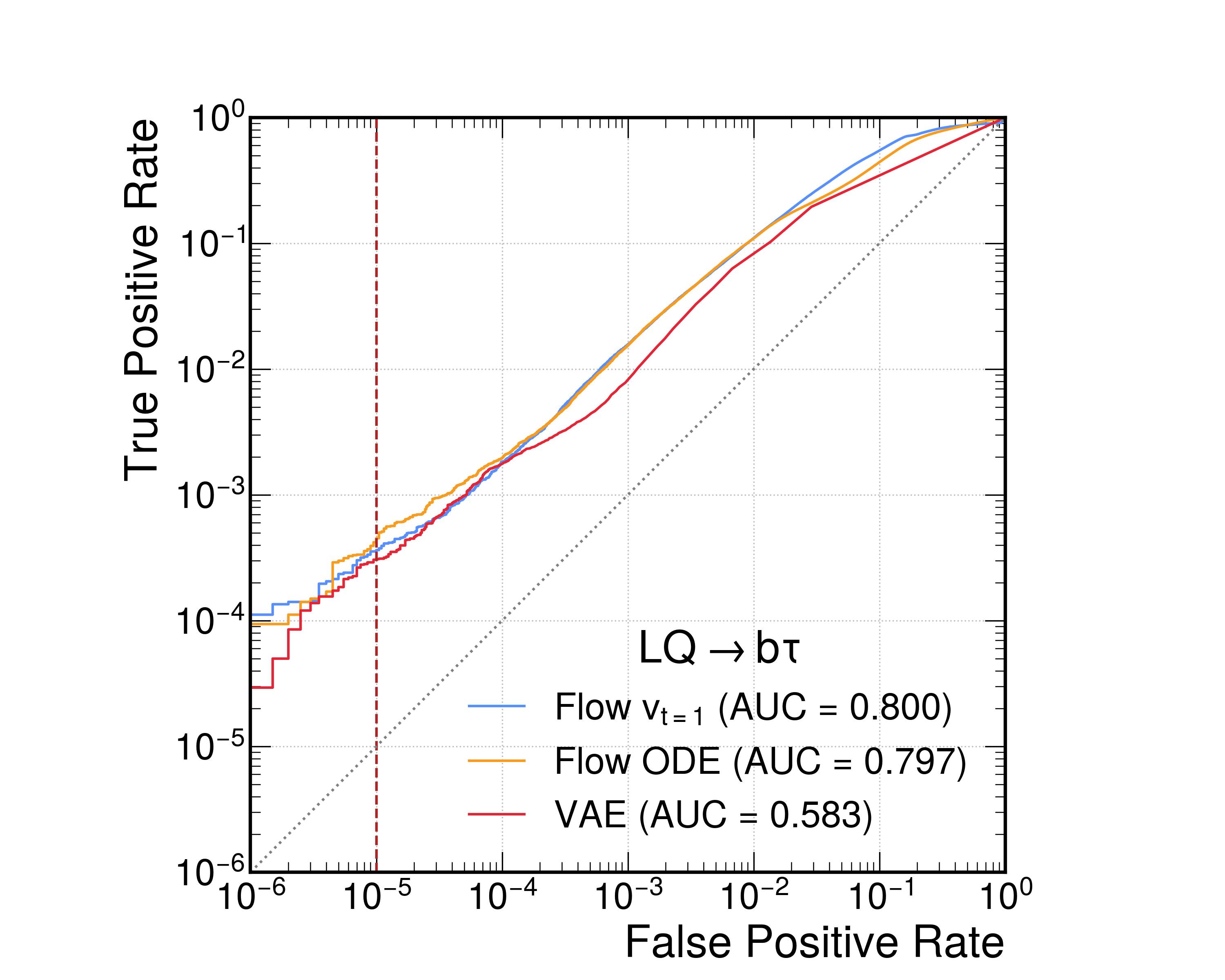}}
    \subfloat
    {\includegraphics[width=.5\textwidth]{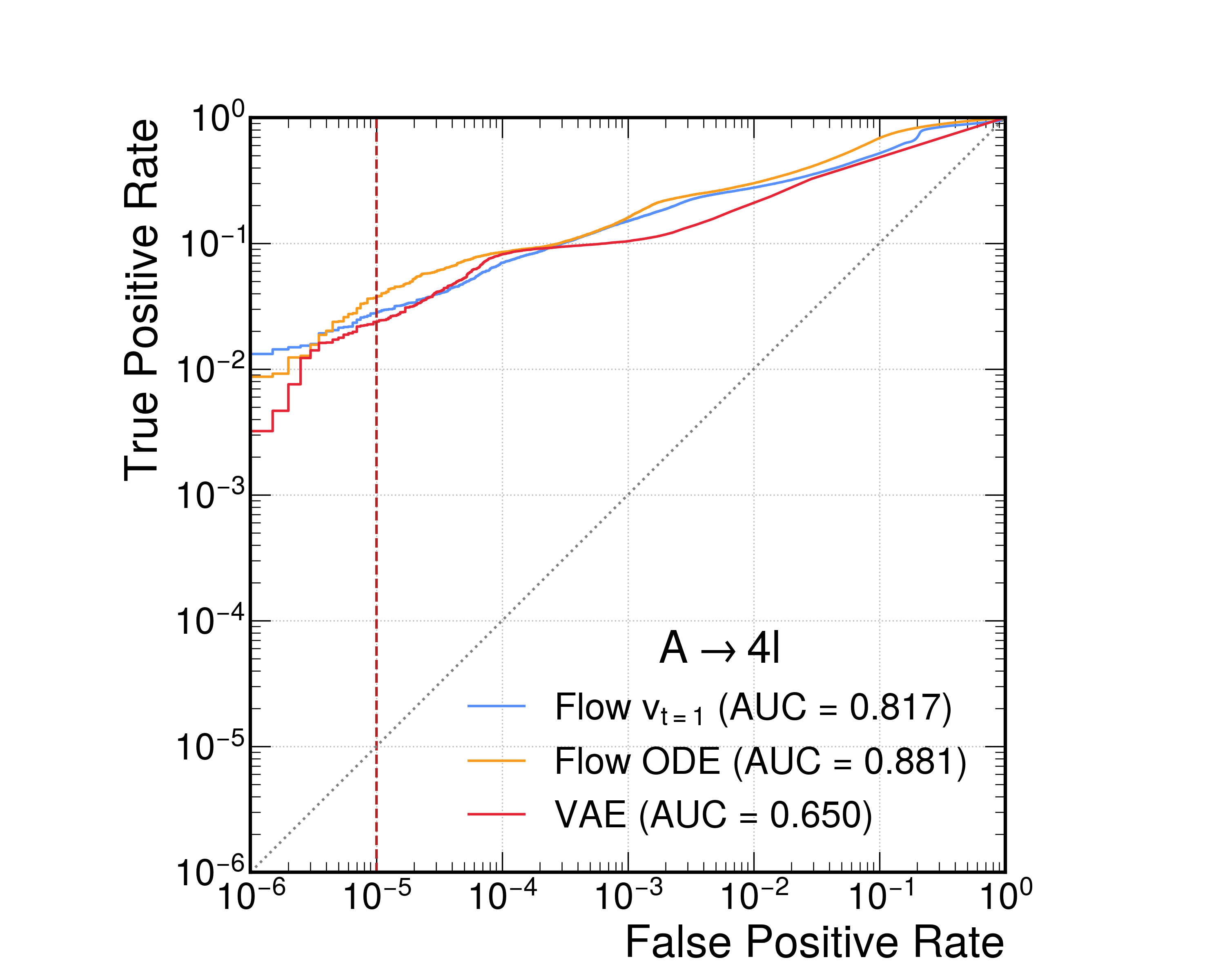}} \\
    \subfloat
    {\includegraphics[width=.5\textwidth]{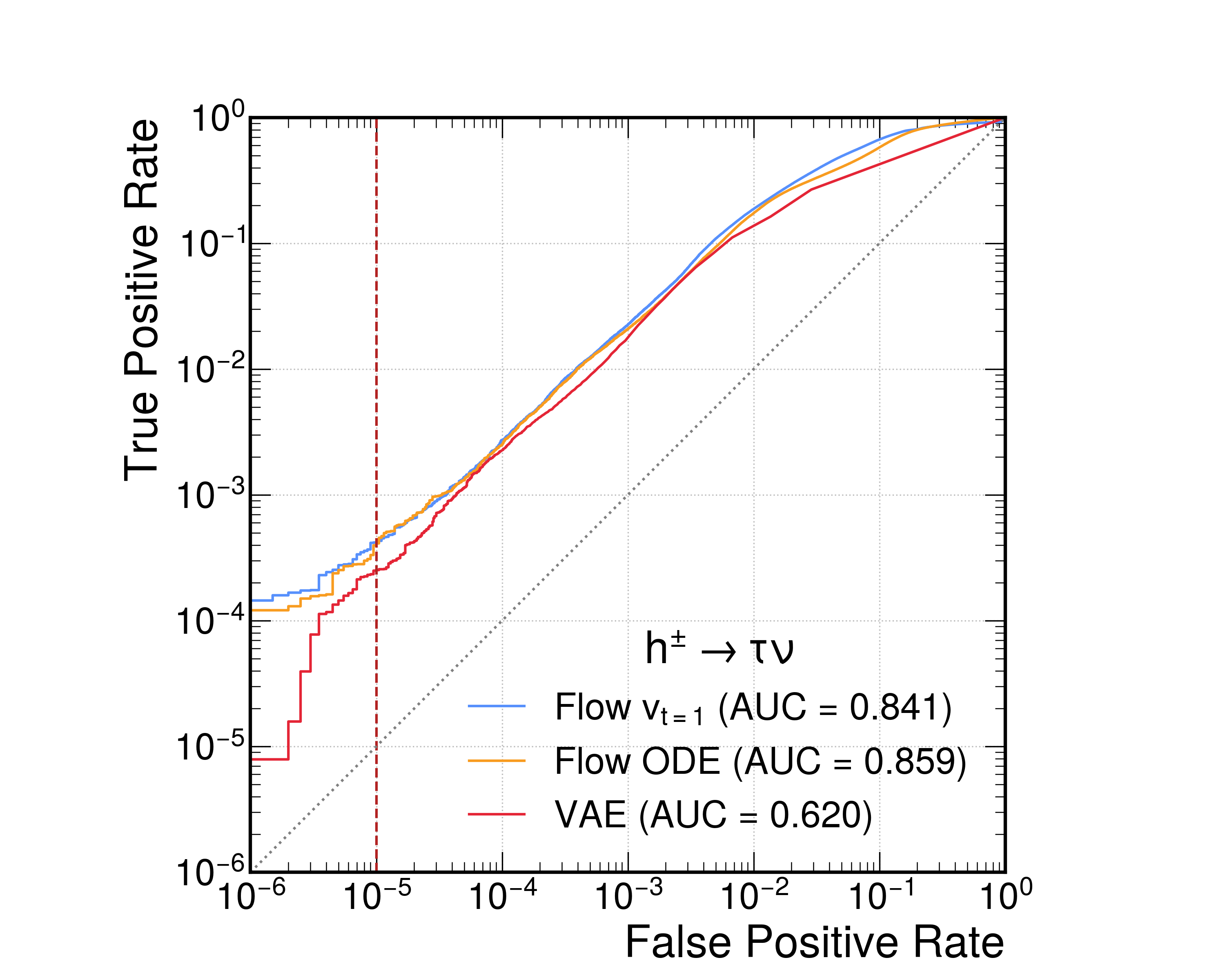}}
    \subfloat
    {\includegraphics[width=.5\textwidth]{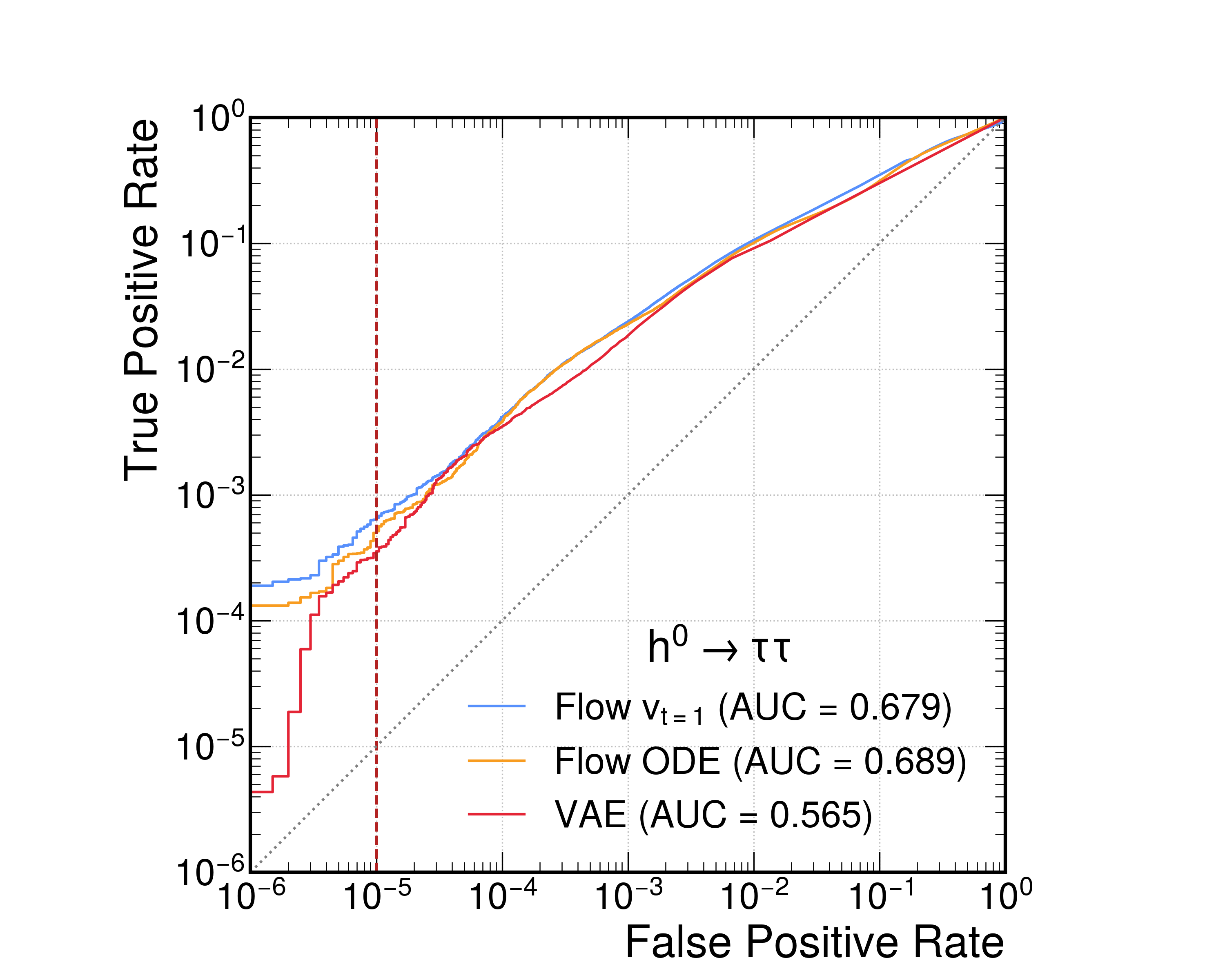}}

    \caption[short]{ROC curves for the 4 anomalous samples. For each sample we show the results for the flow, for both the \emph{Flow} $\mathbf{v}_t$ and \emph{Flow ODE} anomaly scores, and for the VAE. We show as a dashed red line the threshold of $10^{-5}$ FPR at which we evaluate the fraction of TPR for each model; and as a dotted grey line the baseline performance of a random classifier.
        The flow is capable of detecting the anomalies, with both ROC AUC and TPR better than a VAE trained under the same condition.}
    \label{fig:rocs}
\end{figure*}

The receiver operating characteristic
(ROC) curves in Figure \ref{fig:rocs} show the true positive
rate (TPR) as a function of the false positive rate (FPR),
computed by changing the lower threshold applied on the
different anomaly scores. We further quantify the AD
performance quoting the area under the ROC curve (AUC)
and the TPR corresponding to a FPR working point of
$10^{-5}$ (see Table I), which on this dataset corresponds to
the reduction of the background rate to approximately
1000 events per month. We show as a dashed red line the working point of $10^{-5}$ FPR; and as a dotted grey line the baseline performance of a random classifier.

\begin{table*}[]
    \centering
    \caption{Physics performance comparison of different anomaly detection models and scores. Performance is measured by TPR at a fixed FPR of $10^{-5}$ and the Area Under the ROC Curve (AUC).}
    \label{tab:performance}
    \resizebox{2\columnwidth}{!}{%
        \begin{tabular}{l|l|cccc|cccc}
            \hline\hline
            \model                             & \adscore        & \tpr    & \auc                                                                \\
            \cline{3-10}
                                               &                 & \LQbtau & \Allll & \htaunu & \htautau & \LQbtau & \Allll & \htaunu & \htautau \\
            \hline\hline
            VAE from \autocite{Govorkova_2022} & $D_{\text{KL}}$ & 0.07    & 5.27   & 0.08    & 0.11     & 92      & 94     & 94      & 81       \\
            \hline
            \multirow{2}{*}{Flow}              & $\mathbf{v}_t$  & 0.04    & 2.8    & 0.04    & 0.06     & 80      & 82     & 84      & 68       \\
                                               & ODE             & 0.04    & 3.8    & 0.04    & 0.05     & 80      & 88     & 86      & 69       \\
            \hline
            VAE (Ours)                         & $D_{\text{KL}}$ & 0.02    & 2.4    & 0.02    & 0.03     & 58      & 65     & 62      & 57       \\
            \hline\hline
        \end{tabular}
    }
\end{table*}

Table \ref{tab:performance} summarizes the results numerically for both ROC AUC score and TPR, reporting a comparison with the $D_{KL}$ anomaly score from \autocite{Govorkova_2022} as well.

Looking at the results, we see that the flow is capable of detecting anomalies when using either the $\mathbf{v}_t$ or the \emph{ODE} anomaly scores. The \emph{ODE} score has marginally better ROC AUC scores, but similar performances when comparing the TPR values.

Interestingly, the VAE model achieves lower performance than the flow on all processes. 
As already noticed, the VAE performance does not match those shown in \autocite{Govorkova_2022}. This could be due to differences in preprocessing with respect to the published dataset. This does not make our comparison less meaningful, since we used the same preprocessing for both our VAE implementation and the Flow model.

We attribute the differences between the model results in this work and the ones of \autocite{Govorkova_2022} to:
\begin{enumerate}
    \item A difference in the data partitions used for training;
    \item A difference in the exact preprocessing operations used on the data (in which case it would be interesting to assess the performance of the flow under the same preprocessing);
    \item Any remaining difference in the hyperparameters of the NN/training.
\end{enumerate}


\section{Model Compression and HLS C Simulation results} \label{sec6}


As mentioned in section \ref{sec1}, we perform two different compression strategies for porting the model to FPGA.

\subsubsection{PTQ}
We perform some tests and find that a simple, brute force casting of the precision to 18 bits \emph{post-training} is enough for synthesizing the algorithm while having a good tradeoff between performance and resource consumption. The actual weights of the network can be casted down to 12 bits, leaving the 18 bits casting for biases and other operations.

However, we observe that significantly less compression is achievable by default on the \texttt{einsum} operation, responsible for computing the norm squared of the model's output. We find that at least 23 bits are required for casting this operation and retain a good performance, meaning that the majority of resources spent and of the latency will be due to this operation alone.


Performance estimates from the HLS C Simulation of the PTQ model are reported in Table \ref{tab:performancePTQ}. For the same quantization, we report in Table \ref{tab:resources} the resource usage, latency, and initiation interval for the PTQ model deployed on a Xilinx Virtex UltraScale+ FPGA. Resources are based on the Vivado estimates from Vivado HLS.

We observe a good retention of performance, with comparable results to the un-casted model on the TPR and minimal drops of performance on the ROC AUC. The resources being used are a fraction of the available ones on the board, around 7\%, making it possible to host this algorithm on FPGA along with other trigger algorithms, as is usually the case for these applications. The latency of 230 ns and the initiation interval of 5 ns make the algorithm fully compatible with the constraints from the trigger rates of a large LHC experiment and the LHC operational parameters of Run 3.

\subsubsection{QAT}

The model trained with Quantization Aware Training (QAT) is trained using the High Granularity Quantization (HGQ) method \autocite{hgq}.

\begin{figure}[htbp]
    \centering
    \includegraphics[width=0.45\textwidth]{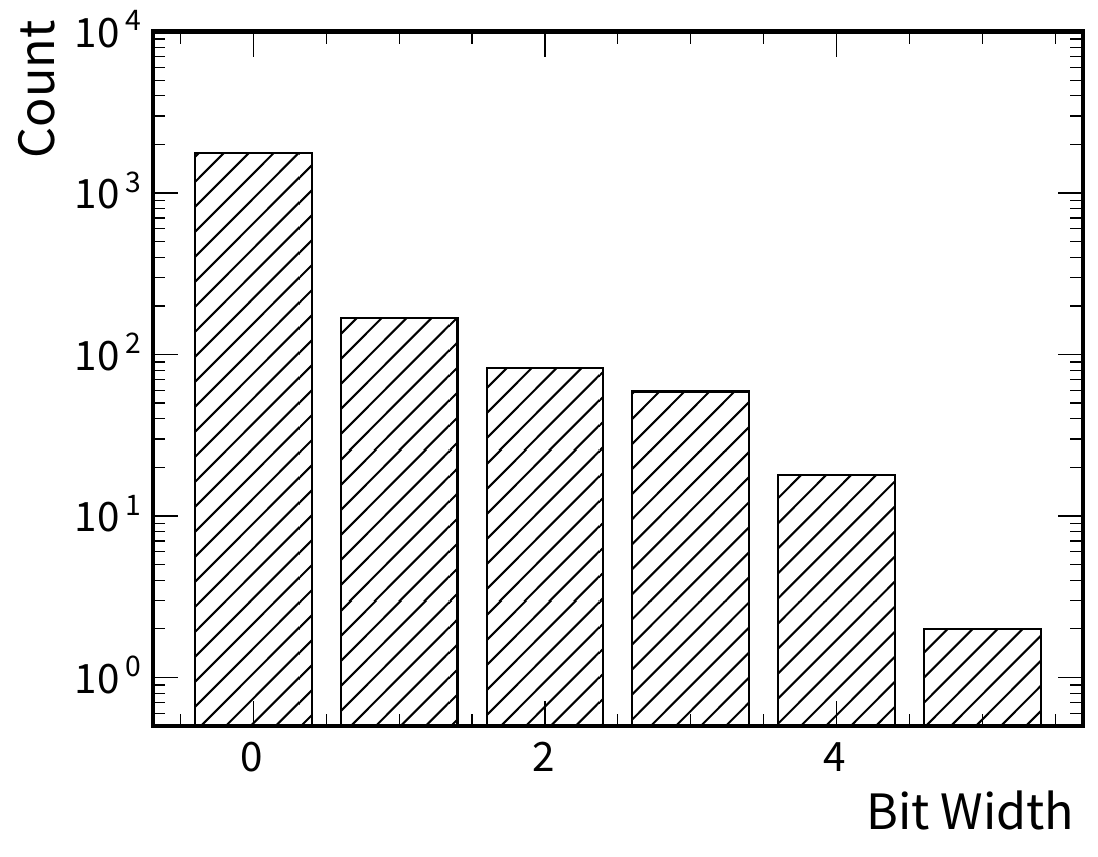}
    \caption{Distribution of bit-widths for the weights of the HGQ-trained model. 84.2\% of the weights are quantized to 0 bits (pruned), while the remaining weights are distributed between 1 and 5 bits.}
    \label{fig:bitdist}
\end{figure}

We initialize the model with 6-fractional bits for weights and activations. We then train the model for 2000 epochs with a cosine decay restarting learning schedule to map out the Pareto front between performance and bit-width with increasing $\beta$ from \texttt{5e-7} to \texttt{5e-6}. Here, $\beta$ is a hyperparameter that controls the trade-off between model performance and bit-width reduction during training, see \autocite{hgq} for details. We show one model obtained with this method in Table \ref{tab:performancePTQ} and Table \ref{tab:resources}. Since HGQ quantizes the weights at a per-weight level and includes zero-bit (pruning), the final model is simultaneously unstructured pruned and quantized. The distribution of bit-widths for the weights of the model is shown in Figure \ref{fig:bitdist}.

We use the da4ml~\cite{da4ml} library to convert the model to RTL design and optimize the constant-matrix-vector multiplication operations (i.e., the dense layers of the MLP) using distributed arithmetic for lower resource usage and latency. Specifically, we convert directly to Verilog, and synthesize with Vivado 2025.1 with a target clock period of 5 ns.

We list in Table \ref{tab:performancePTQ} the physics performance of this approach, while in Table \ref{tab:resources} we report  the usage of resources. We again observe a good retention of performance and a marked improvement in resource usage over the PTQ model, with a latency of just 35\,ns.  All results reported are from RTL behavioral simulation of the RTL designs via Verilator~\cite{verilator}, and all resource usages are from Vivado post place and route reports.

\begin{table*}[]
    \centering
    \caption{Physics performance comparison for the PTQ and HGQ model. Performance is measured by the TPR at a fixed FPR of $10^{-5}$ and the Area Under the ROC Curve (AUC). For the AUC we also report the ratio with the AUC score of the offline Flow $\mathbf{v}_t$ model. }
    \label{tab:performancePTQ}
    \resizebox{2\columnwidth}{!}{%
        \begin{tabular}{l|cccc|cccc}
            \hline\hline
            \model   & \tpr    & \auc                                                                \\
                     & \LQbtau & \Allll & \htaunu & \htautau & \LQbtau & \Allll & \htaunu & \htautau \\
            \hline\hline
            Flow PTQ & 0.03    & 3.6    & 0.04    & 0.06     & 75 (0.94)      & 81 (0.92)     & 81 (0.94)   & 65 (0.94)      \\
            Flow HGQ & 0.04    & 3.4    & 0.05    & 0.06     & 77 (0.96)      & 86 (0.97) & 82 (0.95)    & 66 (0.96)    \\
            \hline\hline
        \end{tabular}
    }
\end{table*}

\begin{table*}[]
    \centering
    \caption{Post-routing resource usage (as total estimates and as percentages over the whole board resources), latency, and initiation interval for the PTQ and HGQ model deployed on a Xilinx Virtex UltraScale+ xcu250-figd2104-2L-e FPGA. The designs are implemented Vivado 2025.1, with a target clock period of 5 ns out-of-context.}
    \label{tab:resources}
    \resizebox{1.7\columnwidth}{!}{%
        \begin{tabular}{l|cccccc}
            \hline\hline
            Model    & DSP           & LUT             & FF              & BRAM & Latency [ns] & II [clk] \\
            \hline \hline
            Flow PTQ & 916 (7.45\%)  & 40,835 (2.36\%) & 11,397 (0.33\%) & 0    & 230         & 1        \\
            Flow HGQ & 28  (<0.01\%) & 5,978  (0.34\%) & 1,683  (0.05\%) & 0    & 35           & 1        \\
            \hline\hline
        \end{tabular}
    }
\end{table*}

\section{Conclusions} \label{sec7}

We have presented the first application of a Continuous Normalizing Flow for unsupervised anomaly detection within the demanding real-time environment of an LHC Level-1 Trigger system. The central innovation of this work is an FPGA-friendly anomaly score, $\mathcal{AS} = \|\mathbf{v}_t\|^2$, derived directly from the model's vector field output, which circumvents the need for computationally expensive ODE integration. We demonstrated that this approach successfully identifies a variety of new physics signatures, achieving performance comparable to a VAE trained under identical conditions. Furthermore, we have shown used advanced quantization techniques to prove that the sythesized algorithm meets the stringent L1T constraints, with a latency ranging from 230 to 35\,ns and minimal resource utilization.

This work opens several avenues for future investigation:
\begin{itemize}
    \item Enhancing the discriminating power of the vector field by incorporating physics-motivated or agnostic constraints into the Flow Matching loss function (Eq.~\ref{eq:loss}).
    \item Exploring alternative anomaly scores derived from the vector field, such as its divergence, and extending the methodology to other ODE-based generative models like Diffusion Models.
    \item Performing a comprehensive scan over model architectures and flow probability paths, to further optimize performance and resource efficiency.
    \item Investigating model pruning and knowledge distillation to potentially simplify the model for inference while preserving its performance.
\end{itemize}

We see Normalizing Flows as not just a useful tool for offline analysis, but also a practical one for deployment in real-time discovery. This study paves the way for integrating this new class of powerful, model-agnostic algorithms into the next generation of trigger systems, expanding the potential for model-agnostic searches at 40\,MHz.

\section*{Acknowledgments}
Francesco Vaselli acknowledges the support and guidance received to kick-off this work in the context of the first CERN Next Generation Triggers hackathon, April 7-11, 2025, CERN, Geneva.

The work done by Dimitrios Danopoulos, Roope Oskari Niemi and Felice Pantaleo is supported by the Eric \& Wendy Schmidt Fund for Strategic Innovation through the CERN Next Generation Triggers project (grant agreement SIF-
2023-004).

Chang Sun is partially supported by the U.S. Department of Energy (DOE), Office of Science, Office of High Energy Physics grant DE-SC0011925, NSF ACCESS Grant PHY240298.

This work was supported by the Open Access Publishing Fund of the Scuola Normale Superiore.

\section*{X1: Code availability} \label{secX1}

The code used for this work is made publicly available under MIT License at the following URL: \href{https://github.com/francesco-vaselli/fAD}{https://github.com/francesco-vaselli/fAD}.

The datasets used are hosted on Zenodo as explained in section \ref{sec2}.




\printbibliography 


\end{document}